\documentclass[%
 reprint,
superscriptaddress, 
 amsmath,amssymb,
 aps,
 prl,
]{revtex4-2}

\newcommand{\supbra}[1]{\left\langle\!\left\langle \, #1 \,\right|\right.}
\newcommand{\supket}[1]{\left.\left| \, #1 \,\right\rangle\!\right\rangle}
\newcommand{\supip}[2]{\left\langle\!\left\langle #1 \middle| #2 \right\rangle\!\right\rangle}

\usepackage{orcidlink}

\usepackage{graphicx}
\usepackage{dcolumn}
\usepackage{bm}
\usepackage{svg}

\usepackage{physics}

\begin{document}


\title{Constrained dissipation and the lifetime of quantum many-body scars}

\author{Weyner Ccuiro~\orcidlink{0000-0001-7572-858X}}
 \email{wccuirom@sissa.it}
\affiliation{International School for Advanced Studies (SISSA), via Bonomea 265, 34136 Trieste, Italy}

\author{Rosario Fazio~\orcidlink{0000-0002-7793-179X}}
\affiliation{%
 The Abdus Salam International Centre for Theoretical Physics, Strada Costiera 11, 34151 Trieste, Italy
}%
\affiliation{Dipartimento di Fisica, Università di Napoli “Federico II”, Monte S. Angelo, I-80126 Napoli, Italy}

\author{Alessandro Silva~\orcidlink{0000-0002-5668-7134}}%
\affiliation{International School for Advanced Studies (SISSA), via Bonomea 265, 34136 Trieste, Italy}

\date{\today}

\begin{abstract}
We investigate the fate of quantum many-body scars in the PXP model subject to Markovian dissipation. We show that when local jump operators respect the same symmetries of the Hamiltonian the full Liouvillian spectrum displays a tower of states with properties analogous to standard quantum scars: large overlaps with Néel states and equally spaced gaps in the imaginary part of the eigenvalues, set by a single frequency. Whether the long-time dynamics is dominated by scars is determined by the real parts of the eigenvalues, which are set by the ratio between the two relaxation rates $\gamma_{\pm}$ describing spin flips down-to-up or up-to-down respectively. For asymmetric dissipation ($\gamma_+ > \gamma_-$) the scarred eigenvalues have the smallest real parts, so that the fidelity oscillations, although always damped with respect to the unitary dynamics, are the longest lived among the dissipative protocols; staggered jump operators further reduce their decay. In contrast, replacing the full projection operator in the dissipators with a partial one breaks the symmetries of the problem and is seen to destroy dissipative scars altogether.


\end{abstract}

\maketitle



The eigenstate thermalization hypothesis (ETH) provides the foundational framework for understanding the emergence of statistical mechanics and thermalization in generic, isolated quantum many-body systems \cite{Rigol2008, gogolin2016, polkovnikov2011, dalessio2016}. While experimental and theoretical milestones have extensively explored equilibration dynamics \cite{kinoshita2006,ueda2020}, the discovery of quantum many-body scars (QMBS) has revealed a robust example of weak ergodicity breaking \cite{Turner2018, Turner2018a, serbyn2020}. 
The archetypal example of this phenomenon is the PXP model, which captures the nearest-neighbor blockade regime of Rydberg atom arrays, a platform successfully deployed in cutting-edge experiments simulating interacting Fibonacci anyons, strongly correlated spin chains, lattice gauge theories \cite{Lesanovsky2012,lesa2012,Surace2019,lin2019}, and bistable phases \cite{SciPostPhys.20.1.019,brighi2026collectivedynamicsspinchains}. Within the PXP model, a measure-zero subspace of non-thermal eigenstates evades thermalization by forming an approximately equally spaced energy tower embedded within the chaotic bulk \cite{Iadecola2019,Iadecola2020,Moudgalya2020}. This distinct algebraic structure allows simple initial configurations, such as the Néel state, to exhibit macroscopic, long-lived periodic revivals rather than rapidly relaxing to equilibrium \cite{Khemani2019,Srednicki1994}. While the phenomenology and systematic construction of QMBS are now well-established in closed systems \cite{Bull2019,Mark2019,PhysRevX4041069}, their fate in the presence of environmental coupling remains a critical frontier.

Quantum scars, being mainly embedded in a dense spectrum of ergodic states, are generically fragile~\cite{Turner2018a,surace2021,surace2023}: perturbations hybridize them with the thermal bulk, although the resulting thermalization can be slow~\cite{Motrunich2020}, and specific deformations can even stabilize them~\cite{michailidis2020stabilizing,Michailidis2019}. This fragility is expected given their conceptual similarity to localized states embedded in a delocalized continuum~\cite{Kirk72}. On the other hand it was recently shown that depending on their timing, measurements can have a positive or detrimental effect on the non-ergodic revival dynamics~\cite{Paviglianiti25}: local measurements of just a single spin, when synchronized with the revivals, can significantly suppress the decay of fidelity oscillations. It is therefore clear that the fate of quantum many-body scars in open environments will depend very much on the specifics: previous studies of quantum many-body scars in open systems have examined their robustness to generic noise such as dephasing~\cite{Ma2025}, identified eigenmodes of the Liouvillian that inherit scar properties~\cite{GarciaGarcia2026}, and shown that suitably engineered dissipation can stabilize scar-dominated steady states or protect scars as dark states~\cite{Wang2024,Jiang2026}. In these works several ingredients, from the type of noise and the timing of measurements to exact symmetries and dark states, act at the same time to stabilize or destabilize the scars, so that a clear physical picture is still missing. When the scars are not protected by an exact symmetry or a decoherence-free subspace, how does the structure of the dissipation itself (the balance between gain and loss, its spatial pattern, and its compatibility with the kinetic constraint) control the lifetime of the revivals?


In this work, we address this question by studying the dynamics of an open PXP model under engineered Markovian dissipation. We consider dissipators that smoothly interpolate between a mechanism of dissipation preserving the fundamental symmetries of the isolated Hamiltonian (Rydberg blockade) and a simple unconstrained local spin-flip. Specifically, we consider a one-dimensional chain of $N$ two-level systems (Rydberg atoms) with periodic boundary conditions described by the Lindbladian $\cal{L}$ defining the master equation  
\begin{equation}
\frac{d\rho}{dt} := \mathcal{L}[\rho] = -i[H_{\text{PXP}}, \rho] + \sum_{\mu} \left( L_\mu \rho L_\mu^\dagger - \frac{1}{2}\{L_\mu^\dagger L_\mu, \rho\} \right).
\end{equation}
In terms of the Pauli matrices $\sigma^z_j$, the interpolation between various regimes of the dynamics is locally achieved by defining the partial projectors $P_{\alpha, j} = 1/(1+\alpha)(1 - \alpha \sigma^z_j)$, which reduce to the identity for $\alpha=0$, $P_{0,j}=\mathbb{I}$, and the full projector for $\alpha=1$, $P_{1,j}=P_{j}=|0\rangle\langle 0|_j$. The coherent dynamics is then defined by the PXP Hamiltonian enforcing a hard constraint (Rydberg blockade) prohibiting simultaneous excitations on nearest-neighbor sites
\begin{equation}
H_{\text{PXP}} = \sum_{j=1}^N P_{j-1}\sigma^x_j P_{j+1}.
\end{equation}
In order to test whether environmental coupling can be structurally engineered to respect the non-ergodic subspace, we construct local jump operators that intrinsically obey the same kinetic constraints as the Hamiltonian:
\begin{equation}
L_{\pm, j} = \sqrt{\gamma_\pm} P_{\alpha,j-1}\sigma^\pm_j P_{\alpha,j+1},
\label{jumpOp}
\end{equation}
giving us the opportunity to study in a single model all aspects of interest (presence/absence of constraints/asymmetries). 
In particular, $\gamma_+$ and $\gamma_-$ represent the gain and loss rates, respectively. Notice that for $\alpha=1$ these dissipators cannot create adjacent excitations, so the dynamics remain strictly confined to the blockaded Hilbert space (the Fibonacci or Lucas cube depending on boundary conditions).
\\
In this framework, we evaluate the full Liouvillian spectrum and find that for weak dissipation the scar tower of the closed system is deformed into Liouvillian scars: the imaginary parts of the eigenvalues closely mirror the closed-system scar towers (leaving the revival period almost invariant and preserving the large-spin precession of the closed-system scars, whose length decays at a rate set by the dissipation \cite{choi2019}) while the real parts, controlling the decay of fidelity oscillations, depend substantially on the asymmetry between gain and loss rates $\gamma_{\pm}$, so that for $\gamma_+>\gamma_-$ the revivals decay much more slowly than for symmetric or reversed rates. The dissipation never raises the revivals above their unitary value: unlike measurements synchronized with the revivals, which can actively rephase the scar dynamics~\cite{Paviglianiti25}, the dissipators considered here can only limit the leakage out of the scar subspace.

Furthermore, we find that substituting uniform dissipation with a spatially staggered application of the jump operators further reduces this leakage. Finally, we probe the fragility of this protective mechanism by relaxing the strict blockade constraint via a partial projection operator ($\alpha<1$). We demonstrate that this explicit kinetic symmetry breaking rapidly kills the revivals upon increasing $\gamma_+$, confirming that the rigid structure of the dissipators is important to constrain decoherence and protect the open-system scars.

\begin{figure}[ht]
    \centering
    \includegraphics[width=\linewidth]{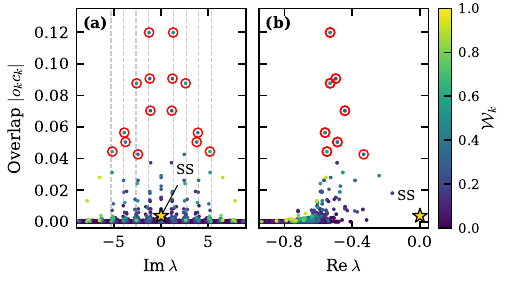}
    \caption{Overlap spectrum $|o_kc_k|$ of the Liouvillian eigenmodes with the N\'eel state [Eq.~(\ref{eq:spectral}) of the Supplemental Material] for $N=14$, $\gamma_+=0.2$, $\gamma_-=0.001$, versus (a) $\mathrm{Im}\,\lambda$ and (b) $\mathrm{Re}\,\lambda$. Each point is coloured by its scar-dyad weight $\mathcal{W}_k$, Eq.~(\ref{eq:Wk}), the overlap of the eigenmode with the $\gamma_\pm=0$ scar subspace. Red circles mark the $N+1$ modes with the largest overlaps and the star the steady state (SS). Dashed lines in (a): bare scar-dyad frequencies $E_m-E_{m'}$ of the closed chain, for the dyads carrying closed-system N\'eel weight $|\langle S_m|\mathbb{Z}_2\rangle|^2|\langle S_{m'}|\mathbb{Z}_2\rangle|^2\geq10^{-2}$. The modes forming the frequency tower are scar-like, carry most of the N\'eel weight and decay more slowly than the bulk of the spectrum, so they govern the transient dynamics.}
    \label{fig:overlap}
\end{figure}

Let us start with the case $\alpha=1$, for which we diagonalize the full Liouvillian of a chain of $N=14$ sites in the two symmetry sectors that contain the N\'eel state (see the Supplemental Material for the symmetries and the numerical details). Exact diagonalization reveals that the defining feature of closed-system scars, the approximately equally spaced tower of eigenstates, persists along the imaginary axis of the Liouvillian spectrum [Fig.~\ref{fig:overlap}(a)]: the eigenmodes with the largest overlap with the N\'eel state sit at $\mathrm{Im}\,\lambda\simeq n\,\Delta E$, with $\Delta E\simeq1.33$ the spacing of the closed-system scar tower, so that the revival period $T=2\pi/\Delta E\simeq4.7$ is essentially unchanged by the dissipation.

To check that these modes are the continuation of the closed-system scars, and not bulk modes that happen to oscillate at the same frequencies, we measure the weight of each eigenmode $\rho_k$ of $\mathcal{L}$ on the scar dyads $|S_m\rangle\langle S_{m'}|$, the eigenmodes of $\mathcal{L}$ at $\gamma_\pm=0$ built from the $N+1$ scar states $|S_m\rangle\propto P_{E_m}|\mathbb{Z}_2\rangle$, i.e.\ the normalized projections of the N\'eel state onto the eigenspaces of $H_{\rm PXP}$ with the largest weight,
\begin{equation}
\mathcal{W}_k=\frac{\sum_{m,m'}\left|\langle S_m|\rho_k|S_{m'}\rangle\right|^2}{\Tr(\rho_k^\dagger\rho_k)}\in[0,1].
\label{eq:Wk}
\end{equation}
This is the colour code of Fig.~\ref{fig:overlap}. For a generic operator $\mathcal{W}_k\simeq3\times10^{-4}$, while the tower modes have $\mathcal{W}_k\simeq0.26$--$0.6$ at $N=14$, reaching $\mathcal{W}_k\simeq0.9$ at the higher harmonics: the open-system tower is a deformation of the closed-system scar dyads. Its scar character decreases with $N$ and is smaller for the other choices of the rates, following the same ordering as the revivals discussed below (see the Supplemental Material, where we also show the spectra for the other rates).

\begin{figure}[ht]
    \centering
    \includegraphics[width=0.7\linewidth]{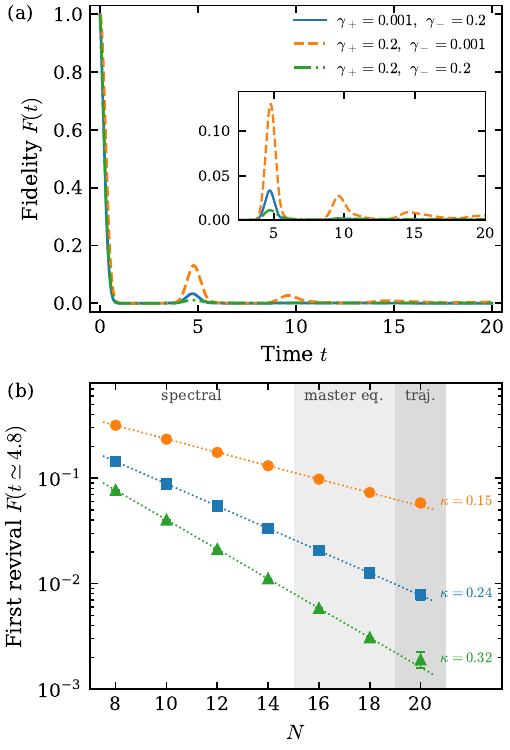}
    \caption{(a) Fidelity $F(t)$ with the N\'eel state for $N=14$ and three choices of the rates, from the Liouvillian spectrum. Inset: zoom on the revivals. Asymmetric rates ($\gamma_+ \neq \gamma_-$) reduce the damping of the revivals relative to the symmetric case, most markedly for $\gamma_+ > \gamma_-$; in all cases the revivals remain well below the unitary ones ($F_1\simeq0.82$ at $N=14$). (b) Height of the first revival ($t\simeq4.8$) versus system size, on a logarithmic scale, with the same colour code as in (a). Values for $N\le14$ are obtained from the Liouvillian spectrum, for $N=16,18$ from the exact integration of the master equation (light shading) and for $N=20$ from $2048$ quantum trajectories (dark shading; error bars are one standard deviation). Dotted lines: fits $F_1\propto e^{-\kappa N}$ over $N=8$--$18$, with the decay rates $\kappa$ indicated.}
    \label{fig:fidelity}
\end{figure}

The frequencies of the scars stay pinned to the bare dyad frequencies $E_m-E_{m'}$ [dashed lines in Fig.~\ref{fig:overlap}(a)] while their decay rates are of order $\gamma$. The reason behind this observation is a symmetry of $\mathcal{L}$ under $\gamma\to-\gamma$. Scale the two rates together, $\gamma_\pm=\gamma\,r_\pm$ at fixed $r_+/r_-$, so that $\mathcal{L}(\gamma)=\mathcal{L}_0+\gamma\mathcal{L}_1$. In the vectorized $\sigma^z$ basis the Hamiltonian part $\mathcal{L}_0=-i(\mathbb{I}\otimes H_{\rm PXP}-H_{\rm PXP}^T\otimes\mathbb{I})$ is purely imaginary, since $H_{\rm PXP}$ is real, while the dissipator $\mathcal{L}_1$ is purely real, since all the jump operators of Eq.~(\ref{jumpOp}) are real matrices. Hence
\begin{equation}
\overline{\mathcal{L}(\gamma)}=-\mathcal{L}(-\gamma),
\label{eq:gammasym}
\end{equation}
for any $N$ and any ratio $\gamma_+/\gamma_-$. If $\lambda(\gamma)$ is an eigenvalue branch emanating from a non-degenerate dyad frequency, $\lambda(0)=-i(E_m-E_{m'})$, Eq.~(\ref{eq:gammasym}) maps it onto itself, $\lambda(-\gamma)=-\overline{\lambda(\gamma)}$. Its decay rate is then odd and its frequency even in $\gamma$,
\begin{equation}
\mathrm{Re}\,\lambda=c_1\gamma+O(\gamma^3),\qquad \mathrm{Im}\,\lambda=-(E_m-E_{m'})+O(\gamma^2),
\end{equation}
where, for real scar states, the first-order rate is
\begin{equation}
c_1\gamma=\sum_{s=\pm,j}\langle S_m|L_{s,j}|S_m\rangle\langle S_{m'}|L_{s,j}|S_{m'}\rangle-\frac{\Gamma_m+\Gamma_{m'}}{2},
\label{eq:c1}
\end{equation}
with $\Gamma_m=\sum_{s=\pm,j}\langle S_m|L_{s,j}^\dagger L_{s,j}|S_m\rangle$. Dissipation thus selects the lifetime of each scar dyad at first order, while its frequency moves only at second order. This is why the Liouvillian tower falls on the dyad energies of the closed system, and why the revival period is robust; the case of quasi-degenerate dyads is discussed in the Supplemental Material.

\begin{figure}[t]
    \centering
    \includegraphics[width=\linewidth]{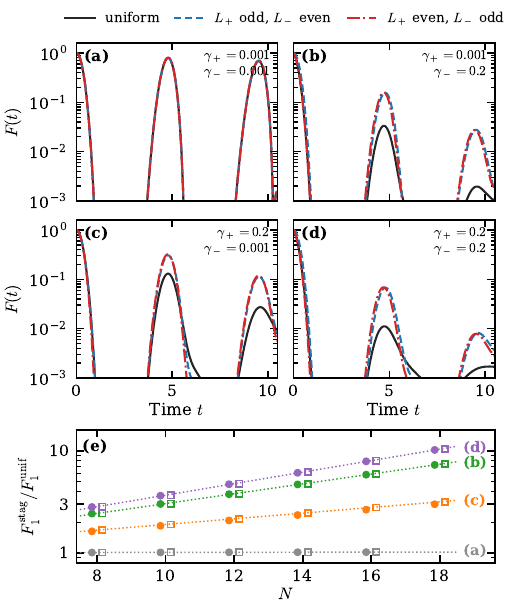}
    \caption{Staggered versus uniform dissipation. (a)--(d) Fidelity $F(t)$ with the N\'eel state $|\mathbb{Z}_2\rangle=|1010\cdots10\rangle$ for $N=14$, for uniform dissipation (solid) and for staggered dissipation with $L_+$ on the odd sites, which are excited in $|\mathbb{Z}_2\rangle$, and $L_-$ on the even ones (dashed), or vice versa (dash-dotted). Staggering has no visible effect for weak dissipation (a), while for larger rates (b)--(d) it markedly reduces the damping of the revivals compared with uniform dissipation. (e) Ratio $F_1^{\rm stag}/F_1^{\rm unif}$ of the first revival heights ($t\simeq4.8$) versus $N$ for the rates of panels (a)--(d) (filled circles: $L_+$ on odd sites; open squares: $L_+$ on even sites; slightly displaced horizontally), with exponential fits to the open squares (dotted). The uniform results coincide with those of Fig.~\ref{fig:fidelity}.}
    
    \label{fig:stag}
\end{figure}

While the bath damps the absolute amplitude of the revivals, their lifetime is strongly regulated by the structure of the dissipators, and asymmetric rates suppress the thermal leakage compared to the symmetric case. Figure~\ref{fig:fidelity}(a) shows the fidelity $F(t)=\Tr[\rho_{\rm Neel}\,\rho(t)]$ for $N=14$, obtained from the spectral decomposition, Eq.~(\ref{eq:spectral}) of the Supplemental Material. The first revival reaches $F\simeq0.13$ for $\gamma_+=0.2$, $\gamma_-=0.001$, against $0.033$ for $\gamma_+=0.001$, $\gamma_-=0.2$ and $0.011$ for $\gamma_+=\gamma_-=0.2$. For comparison, the unitary dynamics gives $F_1\simeq0.82$ at $N=14$ and $0.80$ at $N=16$: even the most favourable rates retain only about $16\%$ and $12\%$ of the closed-system revival. Dissipation therefore never enhances the revivals in absolute terms; what its structure controls is how much of the scar dynamics survives. For larger sizes we obtain $F(t)$ directly from the dynamics, by exact integration of the master equation for $N=16,18$ and by quantum trajectories for $N=20$ [Fig.~\ref{fig:fidelity}(b); see the Supplemental Material]. For every choice of the rates the first revival decays exponentially with the system size, $F_1\propto e^{-\kappa N}$; a fit over $N=8$--$18$ gives $\kappa\simeq0.15$, $0.24$ and $0.32$ for $(\gamma_+,\gamma_-)=(0.2,0.001)$, $(0.001,0.2)$ and $(0.2,0.2)$, and predicts the trajectory results at $N=20$ within one standard deviation. At fixed rates the revivals therefore disappear in the thermodynamic limit, and what the structure of the dissipation controls is the rate $\kappa$. Since $\kappa$ is smallest for $\gamma_+>\gamma_-$, the advantage of the asymmetric choice grows with $N$: at $N=20$ its first revival is about $7$ times larger than for the reversed rates and $30$ times larger than for the symmetric ones. The exponential decay with $N$ is a property of the many-body overlap $F$: local observables such as the staggered magnetization still perform the coherent precession of a large spin, whose length decays at a rate set by the dissipator and does not depend on $N$ (Supplemental Material).

The exponential dependence on $N$ and the ordering of $\kappa$ follow from the first-order rate~(\ref{eq:c1}). The loss term $\Gamma_m$ is a sum of local contributions: $L^\dagger_{+,j}L_{+,j}$ counts the flippable empty sites and $L^\dagger_{-,j}L_{-,j}=\gamma_-n_j$ the excitations, so $\Gamma_m=\gamma_+\langle N_{\rm flip}\rangle_m+\gamma_-\langle N_{\rm exc}\rangle_m$ is extensive and the loss accumulated over one period, $T\Gamma_m$, grows linearly with $N$. The two densities are equal, $\simeq0.276\,N$, for infinite-temperature states, but not for the scars, which, averaging over the whole tower, turn out to have fewer flippable empty sites than excitations, $\langle N_{\rm flip}\rangle\simeq0.23\,N$ against $\langle N_{\rm exc}\rangle\simeq0.335\,N$. Moving rate from $\gamma_-$ to $\gamma_+$ at fixed $\gamma_++\gamma_-$ therefore lowers the loss of the scar dyads and leaves that of the bulk unchanged. For the two cases with the same total rate this predicts $\kappa_{(0.001,0.2)}-\kappa_{(0.2,0.001)}\simeq T\,(0.199)\,(0.335-0.23)\simeq0.10$, in agreement with the measured $0.097$; the jump term in Eq.~(\ref{eq:c1}) shifts the absolute rates but cancels in the difference.

Furthermore, we investigate the spatial configuration of the environmental coupling by replacing the uniform jump operators with a spatially staggered arrangement, in which $L_{+,j}$ acts only on one sub-lattice and $L_{-,j}$ only on the other. We label the sites so that the N\'eel state is $|\mathbb{Z}_2\rangle=|1010\cdots10\rangle$, i.e.\ the odd sites are the excited ones, and consider both placements: $L_+$ on the even sites and $L_-$ on the odd ones, and the reverse. We obtain $F(t)$ by integrating the master equation up to $N=18$ (Supplemental Material). Figure~\ref{fig:stag} shows the result for $N=14$. For weak dissipation [panel (a)] staggering has no visible effect, while for $\gamma_+$ or $\gamma_-$ of order $0.2$ [panels (b)--(d)] it increases the first revival relative to uniform dissipation by a factor $2.4$--$6.2$ at $N=14$ (for $\gamma_+=0.2$, $\gamma_-=0.001$ it reaches $F_1\simeq0.32$, still below the unitary $0.82$). This gain over uniform dissipation is not a small-size effect: the ratio $F_1^{\rm stag}/F_1^{\rm unif}$ grows exponentially with $N$ [Fig.~\ref{fig:stag}(e)], reaching $3.0$--$10.4$ at $N=18$, because the staggered revival decays with a rate $\kappa_{\rm stag}\simeq(0.55$--$0.6)\,\kappa_{\rm unif}$, close to what one obtains by halving the number of active decay channels. The choice of sub-lattice matters only weakly (Supplemental Material).


A simple argument accounts for both observations. In the N\'eel and anti-N\'eel configurations, the turning points of the scar oscillation, every empty site has an excited neighbour, so $L_{+,j}$ is blocked by the constraint and only $L_{-,j}$ acts. With uniform dissipation both turning points decay at rate $N\gamma_-/2$. With staggered dissipation one of the two configurations is exactly dark, since its excitations sit on the sub-lattice where only $L_+$ acts, while the other decays at the same rate $N\gamma_-/2$; the loss at the turning points is therefore halved, independently of which sub-lattice is chosen. Along the rest of the trajectory staggering similarly removes half of the channels available to each process, which explains why the gain over uniform dissipation also appears when $\gamma_+\gg\gamma_-$.

\begin{figure}[ht]
    \centering
    \includegraphics[width=\linewidth]{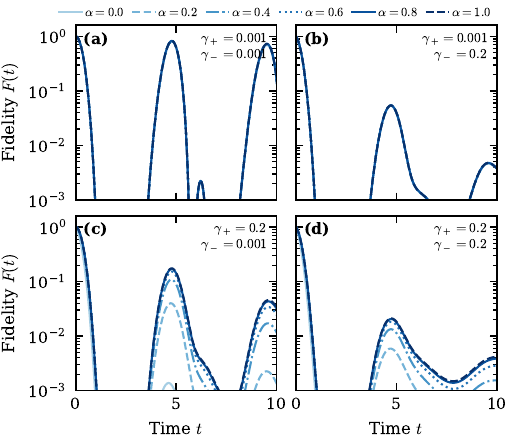}
    \caption{Fidelity $F(t)=\langle \mathbb{Z}_2|\rho(t)|\mathbb{Z}_2\rangle$ (logarithmic scale) for $N=12$ with periodic boundary conditions, when the partial projector $P_\alpha$ replaces the projectors in the jump operators $L_{\pm,j}$, while the PXP Hamiltonian retains the exact blockade. The rates are (a) $\gamma_+=\gamma_-=0.001$, (b) $\gamma_+=0.001$, $\gamma_-=0.2$, (c) $\gamma_+=0.2$, $\gamma_-=0.001$, (d) $\gamma_+=\gamma_-=0.2$. For small $\gamma_+$ [(a), (b)] the curves for different $\alpha$ are indistinguishable, while for $\gamma_+=0.2$ [(c), (d)] lowering $\alpha$ suppresses the revivals by up to two orders of magnitude: in (c) the first revival drops from $F\simeq 0.18$ at $\alpha=1$ to $F\simeq 10^{-3}$ at $\alpha=0$.}
    \label{fig:fig4}
\end{figure}

Finally, we test whether this protection relies on the kinetic constraint by relaxing the strict Rydberg blockade in the dissipators, $0<\alpha<1$. The dissipators then no longer preserve the blockade and the dynamics explores the full $2^N$-dimensional Hilbert space, which limits us to $N=12$ (Supplemental Material). The results are shown in Fig.~\ref{fig:fig4}.
The exact form of the constraint is strictly necessary for this protection. For small $\gamma_+$, modifying $\alpha$ has a negligible effect on the revivals, irrespective of $\gamma_-$, whereas for larger $\gamma_+$ the revivals rapidly deteriorate as $\alpha$ is lowered. Modifying $\alpha$ explicitly breaks the kinetic constraint that supports the scar subspace, but only through the pumping. Since $P_{\alpha,j}|0\rangle_j=|0\rangle_j$, on blockaded configurations $L_{-,j}$, which acts only on an excited site whose neighbours are then necessarily empty, is independent of $\alpha$ and never creates adjacent excitations: for $\gamma_+=0$ the dynamics from the N\'eel state would remain in the constrained subspace for any $\alpha$. On the contrary, for $\alpha<1$, $L_{+,j}$ does create adjacent excitations, which explains the asymmetric sensitivity observed in Fig.~\ref{fig:fig4}.

In conclusion, we explored in a compact setting the effect of the balance between gain and loss, spatial symmetries, and the compatibility of dissipation with the kinetic constraints on the quantum scar tower of the PXP model subject to dissipation.
We demonstrated that the characteristic scar towers persist in the imaginary Liouvillian spectrum, so that the revivals survive the dissipation with an unchanged period. Their damping, which is never compensated as with measurements synchronized with the revivals~\cite{Paviglianiti25}, is controlled by the balance between gain and loss; a staggered application of the dissipators further reduces it, while partial projection breaking the kinetic constraint may rapidly suppress the revivals. 
These results establish a clear framework to extend these ideas to more complex models and generic constructions of scar towers, a goal that is left for future studies. 

 \begin{acknowledgments}
The numerical codes were written and run by WC and tested independently by using AI (Claude by Anthropic).  RF acknowledges financial support through a Google Quantum Research Award and by the European Union (ERC - RAVE, 101053159). WC acknowledges support from the European Union – NextGenerationEU under the National Recovery and Resilience Plan (PNRR) – Mission 4, Component 1, Investment 4.1 (D.M. 118/2023, CUP G93C23000630003).
 \end{acknowledgments}

\bibliography{biblio}

\pagebreak
\widetext
\newpage
\begin{center}
\textbf{\large Supplemental Material}
\end{center}

\setcounter{equation}{0}
\setcounter{figure}{0}
\setcounter{section}{0}

\makeatletter
\renewcommand \thesection{S\@arabic\c@section}
\renewcommand{\theequation}{S\arabic{equation}}
\renewcommand{\thefigure}{S\arabic{figure}}
\renewcommand{\thetable}{S\arabic{table}}

In this Supplemental Material we first describe the exact diagonalization of the Liouvillian: the symmetries used to block-diagonalize it, the spectral decomposition of the fidelity with the N\'eel state, the scar-dyad weight, and the spectra for the choices of the rates not shown in the main text. We then describe the methods used for the time evolution beyond exact diagonalization, and finally test the large-spin picture of the scar dynamics in the presence of dissipation.

\section{Exact diagonalization of the Liouvillian}\label{sec:ed}

\subsection{Symmetry sectors}
For $\alpha=1$ the dynamics is confined to the blockaded configurations, whose number on a ring of $N$ sites is the Lucas number $\mathcal{L}_N$ ($\mathcal{L}_{14}=843$), so that $\mathcal{L}$ acts on a space of dimension $\mathcal{L}_N^2$, $\simeq7\times10^5$ for $N=14$. We reduce it with three linear symmetries of the Liouvillian. Translation, $\mathcal{T}(\rho)=T\rho T^\dagger$, and the site-centred inversion, $\mathcal{I}(\rho)=I\rho I^\dagger$, commute with $\mathcal{L}$ because they leave $H_{\rm PXP}$ invariant and map the jump operators $L_{\pm,j}$ onto $L_{\pm,j+1}$ and $L_{\pm,-j}$, respectively. The third symmetry is
\begin{equation}
\mathcal{S}(\rho)=C\rho^{T}C,\qquad C=\prod_j\sigma^z_j .
\end{equation}
In the $\sigma^z$ basis $H_{\rm PXP}$ is real and anticommutes with $C$, while every jump operator is real and odd under $C$, $CL_{\pm,j}C=-L_{\pm,j}$. In the commutator $-i[H_{\rm PXP},\rho]$ the transposition reverses the order of the factors and $C$ flips the sign of $H_{\rm PXP}$, so that the two signs compensate; in the jump term $L\rho L^\dagger$ the transposition exchanges $L$ and $L^\dagger=L^T$ and the two signs coming from $C$ cancel; finally $L^\dagger L$ is real, symmetric and even under $C$, so the anticommutator is invariant. Hence $\mathcal{S}\mathcal{L}=\mathcal{L}\mathcal{S}$ for any $N$ and any rates. The argument uses only the reality of $H_{\rm PXP}$ and of the jump operators and their parity under $C$, so it holds also for staggered dissipation and for the partial projectors $P_\alpha$, while it would be broken, e.g., by dephasing, $L_j\propto n_j$, or by a longitudinal field.

The N\'eel state $\rho_{\rm Neel}=|\mathbb{Z}_2\rangle\langle\mathbb{Z}_2|$ is invariant under $\mathcal{T}^2$, $\mathcal{I}$ and $\mathcal{S}$. It therefore has weight only in the sectors with momentum $K=0$ and $K=\pi$, inversion parity $p=+1$ and $\mathcal{S}$-parity $\sigma=+1$, exactly $1/2$ in each of them; the unique steady state lies in the sector $(K,p,\sigma)=(0,+,+)$. Within each sector we build an orthonormal basis of Hermitian operators by symmetrizing the dyads $|a\rangle\langle b|$ over the group generated by $\mathcal{T}$, $\mathcal{I}$ and $\mathcal{S}$. Since every Lindbladian preserves Hermiticity, the sector Liouvillian is a real matrix in this basis, which halves the memory needed for its diagonalization. The dimensions of the two sectors are given in Table~\ref{tab:dims}.

\begin{table}[h]
\centering
\begin{tabular}{rrrrr}
\hline\hline
$N$ & $\mathcal{L}_N$ & $\mathcal{L}_N^2$ & $(0,+,+)$ & $(\pi,+,+)$ \\
\hline
8  & 47  & 2\,209   & 109     & 94 \\
10 & 123 & 15\,129  & 478     & 428 \\
12 & 322 & 103\,684 & 2\,418  & 2\,318 \\
14 & 843 & 710\,649 & 13\,347 & 13\,033 \\
\hline\hline
\end{tabular}
\caption{Dimension of the blockaded space, of the full operator space and of the two symmetry sectors $(K,p,\sigma)$ that contain the N\'eel state.}
\label{tab:dims}
\end{table}

We have checked that at $N=8$ the spectra of the sectors coincide with that of the full Liouvillian, and that the fidelity reconstructed from them agrees with the direct propagation of the master equation. At $N=14$ the propagation of the N\'eel state with the two sector Liouvillians reproduces the propagation with the full, unreduced Liouvillian to $5\times10^{-16}$ for $0\le t\le10$, and the N\'eel weights of the eigenmodes sum to one to $5\times10^{-15}$.

\subsection{Fidelity with the N\'eel state from the spectrum}

The Lindblad equation in the vectorization framework has the same form as Schr\"odinger equation
\begin{equation}
    \dv{\rho}{t} = \mathcal{L}[\rho] \quad \Rightarrow \quad \dv{\supket{\rho}}{t} = \mathcal{L}\supket{\rho},
\end{equation}
from where we get a biorthonormal basis, $\mathcal{L}\supket{r_{j}} = \lambda_{j}\supket{r_{j}}$ and $\supbra{l_{j}} \mathcal{L} = \lambda_{j}\supbra{l_{j}}$ with $\supip{l_{i}}{r_{j}}=\delta_{ij}$. 
Since the N\'eel state is pure, the Uhlmann fidelity between the evolved state $\rho(t)$ and $\rho_{\rm Neel}$ simplifies to
\begin{equation}
\begin{split}
    F(t) &= \qty(\Tr[\sqrt{\sqrt{\rho_{\rm Neel}}\rho(t)\sqrt{\rho_{\rm Neel}}}])^{2} = \Tr[\rho_{\rm Neel}\rho(t)]
\end{split}
\end{equation}
which in the vectorized framework reads 
\begin{equation}
    F(t) = \supip{\rho_{\rm Neel}}{\rho(t)},
\end{equation}
this fidelity can be expanded using the biorthonormal basis,
\begin{equation}
\begin{split}
    F(t) &= \supip{\rho_{\rm Neel}}{\rho(t)} = \qty(\sum_{j}o_{j}\supbra{l_{j}}) \qty(e^{\mathcal{L}t}\sum_{k}c_{k}\supket{r_{k}}) = \sum_{k} o_{k} c_{k} e^{\lambda_{k}t},
\end{split}
\label{eq:spectral}
\end{equation}
with the normalization that $1=\Tr{\rho_{\rm Neel}} = \Tr{\rho_{\rm Neel}^2} = \supip{\rho_{\rm Neel}}{\rho_{\rm Neel}} = \sum_{k}o_{k}c_{k}$. Since $\rho_{\rm Neel}$ has weight only in the sectors $(0,+,+)$ and $(\pi,+,+)$, the sum runs over the eigenvalues of these two sectors only.

Given this normalization condition we use the product $o_{k}c_{k}$ to quantify the overlap of the eigenmode with eigenvalue $\lambda_{k}$ with the N\'eel state.

\subsection{Scar-dyad weight}
The closed-system scar states are the normalized projections $|S_m\rangle\propto P_{E_m}|\mathbb{Z}_2\rangle$ of the N\'eel state onto the eigenspaces of $H_{\rm PXP}$, of which we keep the $N+1$ with the largest weight $\|P_{E_m}|\mathbb{Z}_2\rangle\|^2$. At $\gamma_\pm=0$ the eigenmodes of $\mathcal{L}$ are the dyads $|E_a\rangle\langle E_b|$, with eigenvalues $-i(E_a-E_b)$, and the scar-dyad weight $\mathcal{W}_k$, Eq.~(\ref{eq:Wk}) of the main text, measures the weight of the eigenmode $\rho_k$ on the $(N+1)^2$ scar dyads $|S_m\rangle\langle S_{m'}|$. A mode that evolves continuously from a scar dyad as the dissipation is switched on keeps a sizeable $\mathcal{W}_k$, while for a random operator $\mathcal{W}_k\simeq(N+1)^2/\mathcal{L}_N^2$. Note that $\mathcal{W}_k$ measures the weight on the whole scar-dyad space, not only on the dyads with a given frequency. Table~\ref{tab:Wk} gives $\mathcal{W}_k$ for the tower modes with the largest N\'eel overlaps for the rates of Fig.~\ref{fig:overlap} of the main text: they exceed the random value by one to three orders of magnitude for all sizes, although they decrease with $N$.

\begin{table}[h]
\centering
\begin{tabular}{rlll}
\hline\hline
$N$ & $\mathrm{Im}\,\lambda$ of tower modes & $\mathcal{W}_k$ & random \\
\hline
8  & 1.25, 2.63, 3.97, 5.21 & 0.78, 0.95, 0.94, 0.98 & 0.037 \\
10 & 1.28, 2.59, 3.97, 5.25 & 0.73, 0.86, 0.76, 0.93 & 0.008 \\
12 & 1.28, 2.59, 3.94, 5.24, 6.56 & 0.61, 0.72, 0.53, 0.84, 0.93 & 0.0016 \\
14 & 1.28, 1.21, 1.14; 2.61; 3.92, 3.79; 5.19; 6.56 & 0.47, 0.27, 0.26; 0.52; 0.45, 0.36; 0.60; 0.90 & $3.2\times10^{-4}$ \\
\hline\hline
\end{tabular}
\caption{Scar-dyad weight of the tower modes with the largest N\'eel overlaps (positive frequencies), for $\gamma_+=0.2$, $\gamma_-=0.001$, compared with the value for a random operator.}
\label{tab:Wk}
\end{table}

To characterize each harmonic of the revival signal as a whole, we note that the scar energies are nearly equally spaced, $E_m\simeq E_0+m\Omega$, so that the scar dyads with $m-m'=n$ oscillate at $n\Omega$. At finite dissipation we assign an eigenmode to the harmonic $n$ when $\big||\mathrm{Im}\,\lambda_k|-n\Omega\big|<\Omega/2$, with $\Omega\simeq1.26$--$1.30$ fitted from the spectrum, and define the overlap-weighted scar character of the harmonic
\begin{equation}
\overline{\mathcal{W}}_n=\frac{\sum_{k\in n}|o_kc_k|\,\mathcal{W}_k}{\sum_{k\in n}|o_kc_k|}.
\label{eq:Wbar}
\end{equation}
The window contains the deformed scar dyads and any bulk mode at the same frequency, the latter suppressed by their small overlaps.

\paragraph{Frequencies of the tower.} As shown in the main text, the frequency of a non-degenerate scar dyad moves only at second order in the dissipation. For quasi-degenerate dyads the first order can at most split the frequencies symmetrically, leaving the mean frequency of the cluster unchanged; for $\gamma_+=\gamma_-$ the dissipator is symmetric in the dyad basis and there is no first-order splitting at all. Appreciable deviations from the bare dyad frequencies in Fig.~\ref{fig:overlap}(a) of the main text occur only for the tower fragments with the smallest $\mathcal{W}_k$, e.g.\ at $\mathrm{Im}\,\lambda\simeq1.21$ and $1.14$ in the first harmonic (Table~\ref{tab:Wk}), whose bare value is $1.32$; for these modes the hybridization with nearly resonant bulk dyads becomes important.

\subsection{Other choices of the rates}
\begin{figure*}[t]
\centering
\includegraphics[width=0.75\textwidth]{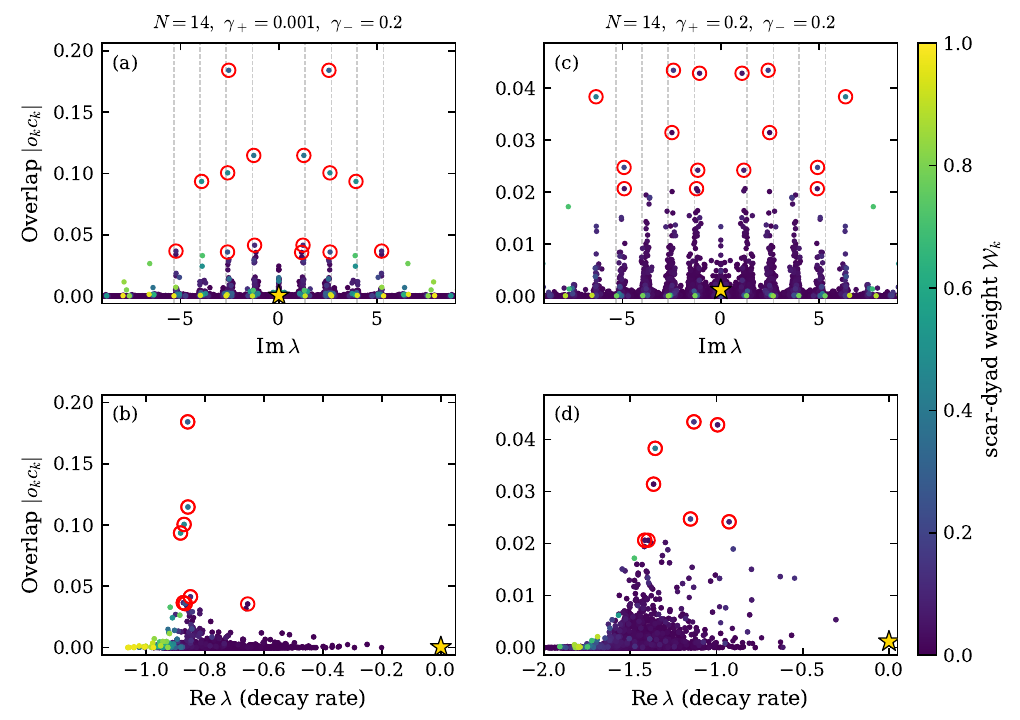}
\caption{Overlap spectra $|o_kc_k|$ of the Liouvillian eigenmodes with the N\'eel state for $N=14$, as in Fig.~\ref{fig:overlap} of the main text, for (a), (b) $\gamma_+=0.001$, $\gamma_-=0.2$ and (c), (d) $\gamma_+=\gamma_-=0.2$, versus $\mathrm{Im}\,\lambda$ (top) and $\mathrm{Re}\,\lambda$ (bottom). Colour: scar-dyad weight $\mathcal{W}_k$; red circles: the $N+1$ modes with the largest overlaps; star: steady state; dashed lines: bare scar-dyad frequencies. Note the different vertical scales and the different ranges of $\mathrm{Re}\,\lambda$.}
\label{fig:overlap_cases}
\end{figure*}

In the main text Fig.~\ref{fig:overlap} shows the overlap spectrum for $\gamma_+=0.2$, $\gamma_-=0.001$. The spectra for the other two choices of the rates considered in the main text are shown in Fig.~\ref{fig:overlap_cases}, and Table~\ref{tab:cases} summarizes the three cases at $N=14$. Notice that the symmetric case has twice the total rate of the other two.

For $\gamma_+=0.001$, $\gamma_-=0.2$ the tower is still clearly visible at the bare dyad frequencies [Fig.~\ref{fig:overlap_cases}(a)], and the largest overlaps are even larger than for $\gamma_+>\gamma_-$, but the tower modes decay faster, $\mathrm{Re}\,\lambda\simeq-0.85$ to $-0.88$ against $-0.33$ to $-0.56$ [compare Fig.~\ref{fig:overlap_cases}(b) with Fig.~\ref{fig:overlap}(b) of the main text], and their scar-dyad weight is smaller. For $\gamma_+=\gamma_-=0.2$ the N\'eel weight is spread over many modes with overlaps below $0.05$, the tower is hardly distinguishable from the bulk, and the modes with the largest overlaps decay with $\mathrm{Re}\,\lambda\simeq-0.9$ to $-1.4$ [Fig.~\ref{fig:overlap_cases}(c),(d)].

\begin{figure*}[t]
\centering
\includegraphics[width=0.85\textwidth]{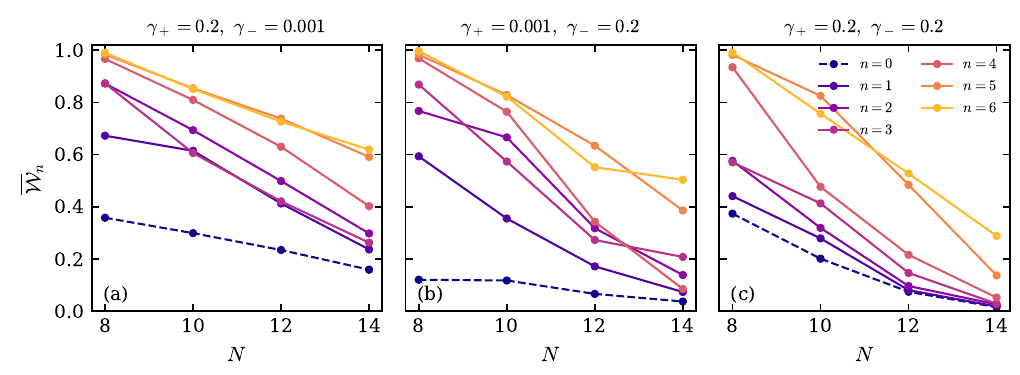}
\caption{Overlap-weighted scar-dyad weight $\overline{\mathcal{W}}_n$, Eq.~(\ref{eq:Wbar}), of the harmonics $n=0,\dots,6$ versus system size, from the Liouvillian spectra for $N=8$--$14$, for the three choices of the rates of the main text. $n=0$ (dashed) collects the non-oscillating modes.}
\label{fig:Wbar}
\end{figure*}

The scar character of each harmonic is shown in Fig.~\ref{fig:Wbar} as a function of $N$. For all the rates $\overline{\mathcal{W}}_n$ decreases with $N$, consistently with the exponential decay of the revivals with system size, while the higher harmonics, less hybridized with the bulk, keep a larger scar character. At $N=14$ the first harmonic has $\overline{\mathcal{W}}_1\simeq0.24$ for $(\gamma_+,\gamma_-)=(0.2,0.001)$, $0.074$ for $(0.001,0.2)$ and $0.018$ for $(0.2,0.2)$: the scar character of the tower follows the same ordering as the revival heights.

\begin{table}[h]
\centering
\begin{tabular}{lcccc}
\hline\hline
$(\gamma_+,\gamma_-)$ & $F_1$ / $F_2$ & $\overline{\mathcal{W}}_1$ & $\overline{\mathcal{W}}_2$ & largest $|o_kc_k|$ \\
\hline
$(0.2,0.001)$ & 0.131 / 0.027  & 0.24  & 0.30  & 0.120 \\
$(0.001,0.2)$ & 0.033 / 0.0020 & 0.074 & 0.139 & 0.184 \\
$(0.2,0.2)$   & 0.011 / 0.0017 & 0.018 & 0.026 & 0.045 \\
\hline\hline
\end{tabular}
\caption{The three choices of the rates at $N=14$: heights of the first two revivals of the fidelity, scar character of the first two harmonics, and largest N\'eel overlap of a decaying mode.}
\label{tab:cases}
\end{table}

\section{Time evolution beyond exact diagonalization}\label{sec:dyn}

Full diagonalization is out of reach beyond $N=14$ on a standard computer, but the revival heights only require $F(t)$, which can be obtained from the dynamics without any spectral information. We either use exact integration of the master equation ($N=16,18$) or quantum trajectories ($N=20$). In the latter case, we compute $F(t)=\mathbb{E}\,|\langle\mathbb{Z}_2|\psi(t)\rangle|^2/\|\psi(t)\|^2$ over waiting-time Monte Carlo trajectories: each state evolves with $H_{\rm eff}=H_{\rm PXP}-\frac{i}{2}\sum_jL_j^\dagger L_j$ until its squared norm falls below a random threshold, after which a jump $L_{\pm,j}$ is applied with probability $\propto\|L_{\pm,j}\psi\|^2$. We use a fourth-order Runge--Kutta scheme with $dt=0.01$ and $2048$ trajectories for each choice of the rates.

The exact integration reproduces the spectral $F(t)$ to $10^{-6}$ at $N=10$ and to $4\times10^{-5}$ at $N=12$, and at $N=16$ halving the time step changes the first revival from $0.0978$ to $0.0980$. The trajectories show no time-step bias at $N=8$, agree with the exact revival heights within $1.3$ standard deviations at $N=10$, and give $0.105\pm0.004$ for the first revival of $(\gamma_+,\gamma_-)=(0.2,0.001)$ at $N=16$, against the exact value $0.098$. The revival heights obtained with the three methods are collected in Table~\ref{tab:revivals}. The fits $F_1\propto e^{-\kappa N}$ over $N=8$--$18$ have residuals below $1\%$ and give $\kappa=0.146$, $0.243$ and $0.322$; their extrapolations to $N=20$, $0.054$, $0.0078$ and $0.0016$, agree with the trajectory results within about one standard deviation. For reference, the unitary dynamics ($\gamma_\pm=0$) gives $F_1=0.821$, $F_2=0.725$ at $N=14$ and $F_1=0.798$, $F_2=0.692$ at $N=16$; the closed-chain revival thus also decreases with $N$, but only with $\kappa_0\simeq0.01$, so that the values of $\kappa$ above are almost entirely due to the dissipation.

\begin{table}[h]
\centering
\begin{tabular}{r ccc ccc}
\hline\hline
 & \multicolumn{3}{c}{first revival ($t\simeq4.8$)} & \multicolumn{3}{c}{second revival ($t\simeq9.6$)} \\
$N$ & $(0.2,0.001)$ & $(0.001,0.2)$ & $(0.2,0.2)$ & $(0.2,0.001)$ & $(0.001,0.2)$ & $(0.2,0.2)$ \\
\hline
8  & 0.317 & 0.144 & 0.0770 & 0.128 & 0.0292 & 0.0266 \\
10 & 0.234 & 0.0882 & 0.0403 & 0.0769 & 0.0119 & 0.0106 \\
12 & 0.175 & 0.0542 & 0.0212 & 0.0458 & 0.00478 & 0.00423 \\
14 & 0.131 & 0.0334 & 0.0111 & 0.0273 & 0.00195 & 0.00170 \\
16 & 0.0978 & 0.0205 & 0.00585 & 0.0163 & 0.00080 & 0.00068 \\
18 & 0.0731 & 0.0126 & 0.00308 & 0.00977 & 0.00033 & 0.00027 \\
20 & $0.058(3)$ & $0.0079(8)$ & $0.0019(3)$ & $0.0063(7)$ & $0.00012(3)$ & $0.00014(3)$ \\
\hline\hline
\end{tabular}
\caption{Revival heights of the fidelity for $(\gamma_+,\gamma_-)$ as indicated: $N\le14$ from the Liouvillian spectrum, $N=16,18$ from the exact integration of the master equation, $N=20$ from $2048$ quantum trajectories (statistical error on the last digit in brackets).}
\label{tab:revivals}
\end{table}

Since the staggered jump operators still have the form $P\sigma^\pm P$, the dynamics remains in the constrained space, and we obtain $F(t)$ with the same exact integration, with site-dependent rates, up to $N=18$ ($dt=0.04$ up to $t=10.5$ for $N\le16$, $dt=0.05$ up to $t=6.5$ for $N=18$). The uniform results obtained in this way coincide with those of Table~\ref{tab:revivals} within $0.2\%$. The choice of sub-lattice matters only weakly: placing $L_+$ on the even sites, so that $L_-$ acts on the sites excited in $|\mathbb{Z}_2\rangle$, is slightly better for all the rates we considered, by $1$--$2\%$ for $\gamma_-\gg\gamma_+$ and by about $5\%$ for $\gamma_+\gg\gamma_-$. The argument of the main text, which explains the gain over uniform dissipation by the halving of the loss at the turning points of the oscillation, is independent of the placement; the small difference comes from the intermediate configurations visited during the oscillation, which are not symmetric between the two sub-lattices.

For $\alpha<1$ the operators $L_{+,j}$ create adjacent excitations when acting on blockaded configurations, so the dynamics explores the full $2^N$-dimensional Hilbert space. We integrate the master equation for the dense $\rho$ in this space with a fourth-order Runge--Kutta scheme ($dt=0.04$), which limits us to $N=12$.

\section{Large-spin dynamics under dissipation}\label{sec:spin}

In the closed PXP chain the revivals from the N\'eel state are commonly interpreted as the precession of a large collective spin, the scar states forming an approximate spin-$N/2$ multiplet of an emergent SU(2) algebra~\cite{choi2019}. Here we check whether this picture survives the dissipation, using observables that, unlike the fidelity, are averages of local operators. We consider the staggered magnetization and its conjugate,
\begin{equation}
Z=\frac1N\sum_j(-1)^j\sigma^z_j,\qquad Y=\frac1N\sum_j(-1)^jP_{j-1}\sigma^y_jP_{j+1},
\end{equation}
with $\sigma^z|1\rangle=|1\rangle$, so that $\langle Z\rangle=1$ in the N\'eel state $|\mathbb{Z}_2\rangle=|1010\cdots10\rangle$. Under the PXP Hamiltonian alone one has $\frac{d}{dt}\langle Z\rangle=2\langle Y\rangle$, so that a precession at the scar frequency $\Omega$ about the axis generated by $H_{\rm PXP}$ traces a circle in the plane $(\langle Z\rangle,2\langle Y\rangle/\Omega)$, whose radius is the length of the collective spin normalized to its initial value. We therefore define the spin length
\begin{equation}
\ell(t)=\sqrt{\langle Z\rangle_t^2+\left(2\langle Y\rangle_t/\Omega\right)^2},
\end{equation}
with $\Omega=2\pi/T$ and $T$ the time of the first revival of the closed chain ($\Omega\simeq1.31$ for $N=14$ and $16$). The third component, $\langle H_{\rm PXP}\rangle/N$, vanishes identically at all times: the initial state is invariant under the symmetry $\rho\mapsto C\rho^TC$ of the Liouvillian, which maps $\langle H_{\rm PXP}\rangle$ into $-\langle H_{\rm PXP}\rangle$, so the precession remains confined to the $(Z,Y)$ plane also in the presence of dissipation.

We obtain $\rho(t)$ by integrating the master equation directly in the constrained space with a fourth-order Runge--Kutta scheme ($dt=0.05$), for the closed chain and for the three choices of rates of Fig.~\ref{fig:fidelity}, at $N=14$ and $16$. The trace is conserved to $10^{-14}$, and the fidelity obtained in the same runs reproduces the first revivals of Fig.~\ref{fig:fidelity} [e.g.\ $F_1=0.0978$, $0.0205$ and $0.00585$ at $N=16$]. The code was checked at $N=6$ against an independent integration of the Lindblad equation in the full $2^N$-dimensional Hilbert space, with agreement to $2\times10^{-8}$.

\begin{figure}[h]
\centering
\includegraphics[width=0.45\textwidth]{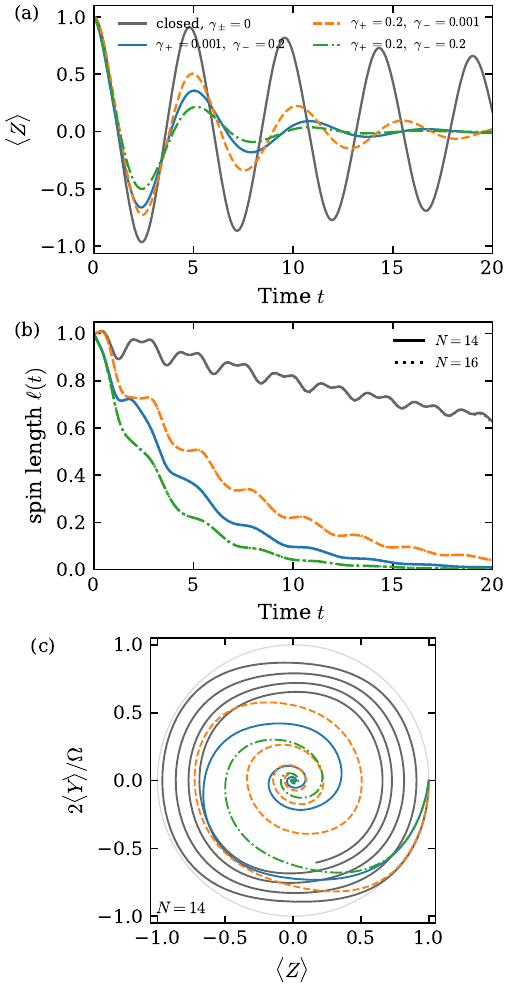}
\caption{Large-spin dynamics from the N\'eel state for the closed chain and the three choices of rates of Fig.~\ref{fig:fidelity}. (a) Staggered magnetization $\langle Z\rangle$ for $N=14$. (b) Spin length $\ell(t)$ for $N=14$ (lines as in panel a) and $N=16$ (dotted); the two sizes are indistinguishable on this scale. (c) Trajectory in the plane $(\langle Z\rangle,2\langle Y\rangle/\Omega)$ for $N=14$; the grey circle has unit radius.}
\label{fig:spin}
\end{figure}

The results are shown in Fig.~\ref{fig:spin}. In all cases the trajectory in the $(\langle Z\rangle,2\langle Y\rangle/\Omega)$ plane is a smooth inward spiral at the scar frequency [Fig.~\ref{fig:spin}(c)]: the dissipation shrinks the collective spin but does not destroy the coherence of its precession. The spin length [Fig.~\ref{fig:spin}(b) and Table~\ref{tab:spin}] decays in the same order as the fidelity, slowest for $\gamma_+>\gamma_-$ and fastest for symmetric rates, while in the closed chain it decreases only slowly, reflecting the approximate nature of the SU(2) algebra of the PXP scars. In contrast to the first revival of the fidelity, which decreases exponentially with $N$ [Fig.~\ref{fig:fidelity}(b)], $\ell(t)$ coincides for $N=14$ and $16$ within $1\%$ (and within $3\%$ of the values we find for $N=8$ and $10$). The exponential dependence of $F$ on system size thus reflects the many-body nature of the overlap with the N\'eel state, while the amplitude of the oscillations of local observables, the quantity accessible in Rydberg-atom experiments, is independent of $N$ and is controlled by the structure of the dissipation in the same way as the revivals of the fidelity.

\begin{table}[h]
\centering
\begin{tabular}{lcccc}
\hline\hline
 & \multicolumn{2}{c}{$\ell$ at $t\simeq4.8$} & \multicolumn{2}{c}{$\ell$ at $t\simeq9.6$} \\
$(\gamma_+,\gamma_-)$ & $N=14$ & $N=16$ & $N=14$ & $N=16$ \\
\hline
closed & 0.909 & 0.909 & 0.818 & 0.816 \\
$(0.2,0.001)$ & 0.504 & 0.504 & 0.221 & 0.221 \\
$(0.001,0.2)$ & 0.374 & 0.374 & 0.103 & 0.102 \\
$(0.2,0.2)$ & 0.224 & 0.224 & 0.051 & 0.051 \\
\hline\hline
\end{tabular}
\caption{Spin length at the times of the first two revivals of the closed chain.}
\label{tab:spin}
\end{table}

\end{document}